\documentclass[aps,noshowpacs,floatfix,superscriptaddress,preprintnumbers,showpacs,showkeys,nofootinbib]{revtex4}
\usepackage{graphicx}
\usepackage{amsmath}
\usepackage{subfigure}
\usepackage{slashed,braket}
\usepackage{bbold}

\newcommand{\be}{\begin{equation}}
\newcommand{\ee}{\end{equation}}
\newcommand{\ba}{\begin{eqnarray}}
\newcommand{\ea}{\end{eqnarray}}
\newcommand{\nn}{\nonumber}

\newcommand{\eq}{\begin{eqnarray}}
\newcommand{\en}{\end{eqnarray}}

\begin{document}

\title{Relating CP-violating decays to the neutron EDM}

\author{Thomas~Gutsche}
\affiliation{Institut f\"ur Theoretische Physik, 
Universit\"at T\"ubingen, 
Kepler Center for Astro and Particle Physics, 
Auf der Morgenstelle 14, D-72076 T\"ubingen, Germany} 
\author{Astrid~N.~Hiller Blin} 
\affiliation{
              Institut f\"ur Kernphysik \& PRISMA Cluster of Excellence, Johannes Gutenberg Universit\"at, D-55099 Mainz, Germany}  
\author{Sergey~Kovalenko}
\affiliation{Departamento de F\'\i sica y Centro Cient\'\i fico 
Tecnol\'ogico de Valpara\'\i so (CCTVal), Universidad T\'ecnica 
Federico Santa Mar\'\i a, Casilla 110-V, Valpara\'\i so, Chile}
\author{Serguei~Kuleshov}
\affiliation{Departamento de F\'\i sica y Centro Cient\'\i fico 
Tecnol\'ogico de Valpara\'\i so (CCTVal), Universidad T\'ecnica 
Federico Santa Mar\'\i a, Casilla 110-V, Valpara\'\i so, Chile} 
\author{Valery~E.~Lyubovitskij} 
\affiliation{Institut f\"ur Theoretische Physik, Universit\"at T\"ubingen, 
Kepler Center for Astro and Particle Physics, 
Auf der Morgenstelle 14, D-72076 T\"ubingen, Germany}
\affiliation{Departamento de F\'\i sica y Centro Cient\'\i fico 
Tecnol\'ogico de Valpara\'\i so (CCTVal), Universidad T\'ecnica 
Federico Santa Mar\'\i a, Casilla 110-V, Valpara\'\i so, Chile} 
\affiliation{Department of Physics, Tomsk State University, 
634050 Tomsk, Russia} 
\affiliation{Laboratory of Particle Physics, Mathematical Physics Department,
Tomsk Polytechnic University, 634050 Tomsk, Russia}
\author{Manuel J.~Vicente Vacas}
\affiliation{Instituto de F\'{\i}sica Corpuscular, 
Universidad de Valencia--CSIC,\\
Institutos de Investigaci\'on, Ap. Correos 22085, E-46071 Valencia, Spain} 
\author{Alexey~Zhevlakov} 
\affiliation{Department of Physics, Tomsk State University, 
634050 Tomsk, Russia}

\today

\begin{abstract}
We use the present upper bound on the neutron electric dipole moment to give an estimate for the upper limit of the CP-violating couplings of the $\eta(\eta')$ to the neutron. Using this result, we derive constraints on the CP-violating two-pion decays of the $\eta(\eta')$ mesons. Our results are relevant for the running and planned GlueX and LHCb measurements of rare meson decays.
\end{abstract}

\keywords{Neutron EDM, CP-violating decays}
\maketitle

\section{Introduction}
\label{SEDMIntro}

The understanding of CP-violating processes is important from many points of view, both in particle physics, as well as for hadron interactions. On the one hand, the baryon asymmetry in the universe is tightly connected to the existence of CP-violation. In the standard model (SM), this violation has already been confirmed experimentally~\cite{Agashe:2014kda}, and it is explained by the explicit symmetry breaking due to a complex phase in the Cabibbo-Kobayashi-Maskawa matrix. However, it is insufficient by several orders of magnitude to explain the size of the baryon asymmetry observed. One expects further beyond-the-SM CP-violating processes to enhance this effect~\cite{Kuckei:2005pg,Pospelov:2005pr,Faessler:2006at,Dib:2006hk,Faessler:2006vi}.

On the other hand, the existence of a permanent electric dipole moment (EDM), both in elementary particles and in hadrons such as the neutron, would point to CP violation as well~\cite{Pich:1991fq,Dar:2000tn,Gorchtein:2008pe,Jarlskog:1995ww,Peccei:1977hh,Peccei:1977ur}. The neutron, in particular, is a favoured target due to its neutral charge and stability. Currently, only an upper bound on its EDM is known,  $|d_{n}| \leq 2.9\times 10^{-26}e$~cm with a confidence level (C.L.) of $ 90\%$~\cite{Agashe:2014kda}.

In view of the extensive experimental programs, by collaborations such as LHCb and GlueX~\cite{Aaij:2016jaa,Ghoul:2015ifw}, aiming to measure EDMs and searching for rare CP-violating decays, we investigate how to gain further theoretical insight on the size of CP-violating decays, and how to relate them to the permanent EDM of the neutron. We use the framework of fully covariant chiral perturbation theory (ChPT)~\cite{Weinberg:1978kz,Gasser:1987rb}, with the explicit inclusion of vector-meson (VM)~\cite{Borasoy:1995ds,Drechsel:1998hk,Kubis:2000aa,Schindler:2005ke} and $\Delta(1232)$~\cite{Pascalutsa:2000kd} degrees of freedom. Furthermore, the renormalization scheme used is the extended on-mass shell (EOMS) scheme~\cite{Gegelia:1999gf,Fuchs:2003qc}, which is relativistic, satisfies analyticity, and usually 
converges faster than other relativistic or non-relativistic approaches.

We focus on the process $\eta (\eta') \rightarrow\pi\pi$. Since it contributes to the neutron EDM, the above-cited current experimental limits on the latter allow for the derivation of indirect upper bounds on the branching ratios of this CP-violating type of decays~\cite{Gutsche:2016jap}. The current direct experimental 90\% C.L. upper 
limits~\cite{Agashe:2014kda,Aaij:2016jaa} are
\begin{align}
\nn\frac{\Gamma(\eta\rightarrow\pi\pi)}{\Gamma_\eta^{\text{tot}}}
&<
\Bigg\{\begin{array}{c}
1.3\times 10^{-5}\text{ for }\pi^+\pi^-\\
3.5\times 10^{-4}\text{ for }\pi^0\pi^0
\end{array},\\
\frac{\Gamma(\eta'\rightarrow\pi\pi)}{\Gamma_{\eta'}^{\text{tot}}}
&<
\Bigg\{\begin{array}{c}
1.8\times 10^{-5}\text{ for }\pi^+\pi^-\\
4.0\times 10^{-4}\text{ for }\pi^0\pi^0
\end{array}\,,\label{Epdgbranch}
\end{align}
with $\Gamma_{\eta(\eta')}^{\text{tot}}$ the total decay width.

We proceed as follows: with the help of the experimental upper limit on the neutron EDM, we derive a constraint on the CP-violating couplings of the $\eta(\eta')$ to the neutron. By describing these couplings via an intermediate pion loop, we can thus extract an estimate of the upper limit of the CP-violating $\eta(\eta')$ decay into two pions, which turns out to be much more stringent than those given by the experiment so far.

\section{Relating the neutron EDM to the CP-violating couplings}\label{STheoetaNN}

\begin{figure}[htbp]
\begin{center} 
\subfigure[]{ 
\label{fedma} 
\includegraphics[width=0.3\textwidth]{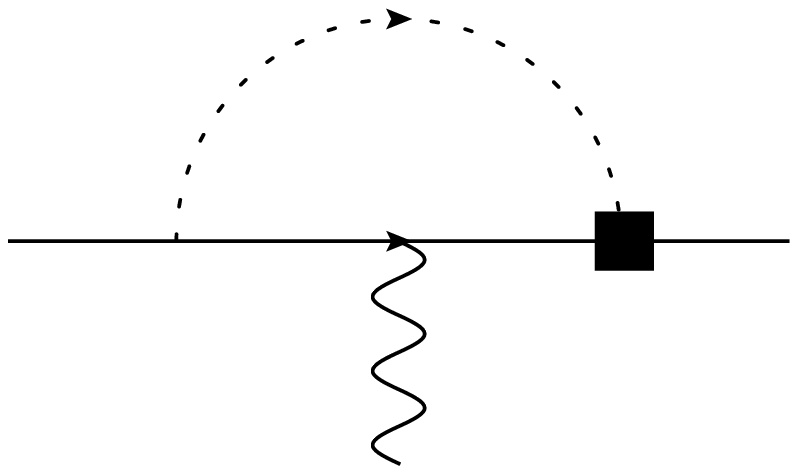}\hspace{2cm}
\includegraphics[width=0.3\textwidth]{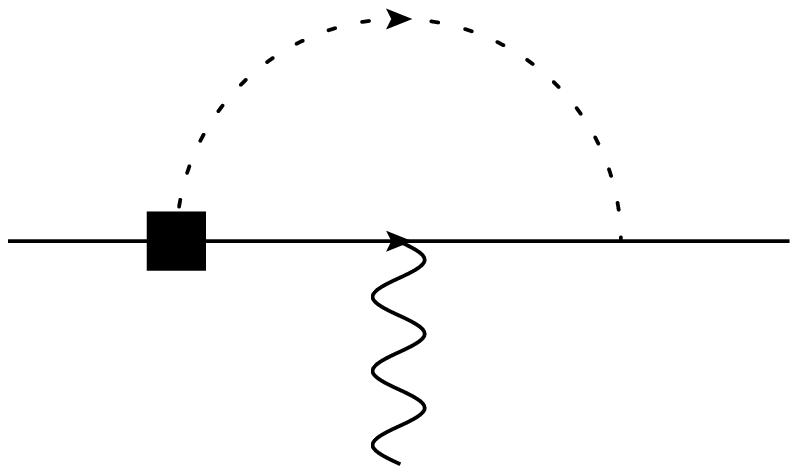}}\\ 
\subfigure[]{ 
\label{fedmb} 
\includegraphics[width=0.3\textwidth]{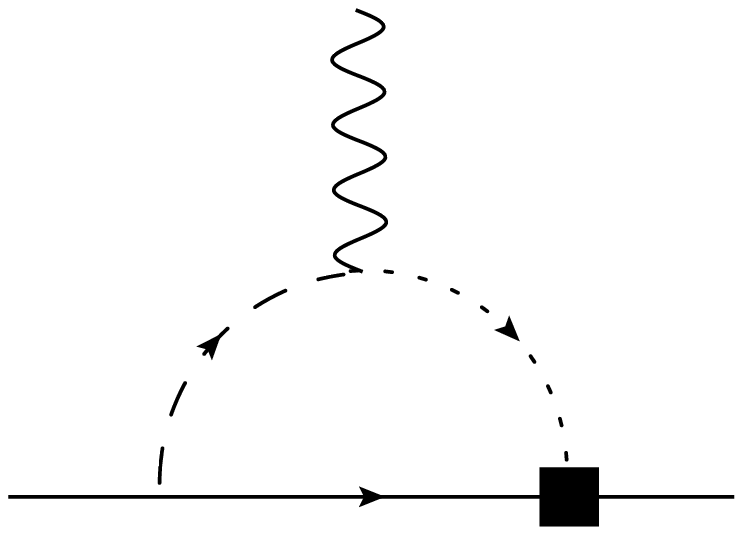}\hspace{2cm} 
\includegraphics[width=0.3\textwidth]{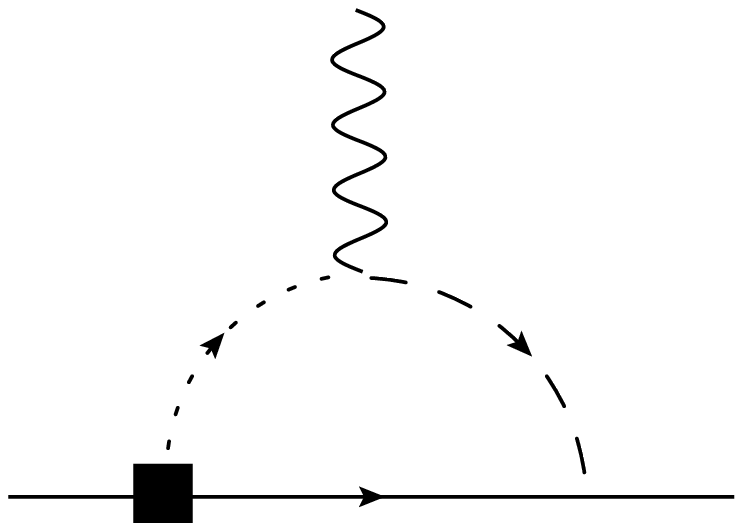}} 
\end{center} 
\caption{Loops that can contribute to the neutron EDM. 
The solid line represents the neutron, the dotted line the $\eta(\eta')$, 
the dashed lines are vector-meson contributions, and the wavy line 
corresponds 
to the photon. Again, the black box stands for a $CP$-violating vertex.}
\label{fedm}
\end{figure}

The neutron EDM is extracted from the CP-violating piece of the vector current $J^\mu$ of a virtual photon coupling to the neutron, which reads: 
\begin{align}
\bra{B(p')}J^\mu_\text{CPV}\ket{B(p)} 
=\bar{u}(p')\frac{\mathrm{i}\sigma^{\mu\nu}\gamma_5q_\nu}{4m} 
F_\text{EDM}(q^2)u(p)\,,
\end{align} 
where $q_\nu$ is the momentum of the virtual photon, $\epsilon_\mu$ 
its polarization, and 
$\sigma^{\mu\nu}=\frac{\mathrm{i}}{2}\,[\gamma^{\nu},\gamma^{\nu}]$. 
At the point where $q^2=0$, 
the form factor reduces to the electric dipole moment 
$F_\text{EDM}(0) = \tilde{d}_N$. 
In our model, we consider the contributions to the CP violation that arise through the loops of Fig.~\ref{fedm}. 

The standard ChPT Lagrangians that describe the CP-conserving couplings appearing 
here are detailed in Ref.~\cite{Gutsche:2016jap}. These include all the couplings involving photons, pseudoscalar and vector mesons, nucleons, and the $\Delta(1232)$ resonance. As for the $CP$-violating coupling of the $\eta(\eta')$ to the neutron, it is given by
\begin{align}
\mathcal{L}_{HNN}^{CP} =g^{CP}_{HNN} \, H \, \bar{N}N\,,
\end{align}
with $H=\eta,\eta'$. The unknown couplings $g^{CP}_{HNN}$ can thus be related to the size of the neutron EDM $F_\text{EDM}(0)$, thus leading to a constraint on their size.

\begin{figure}[htbp]
\begin{center}
\subfigure[]{ 
\label{fetaNloopa} 
\includegraphics[width=0.3\textwidth]{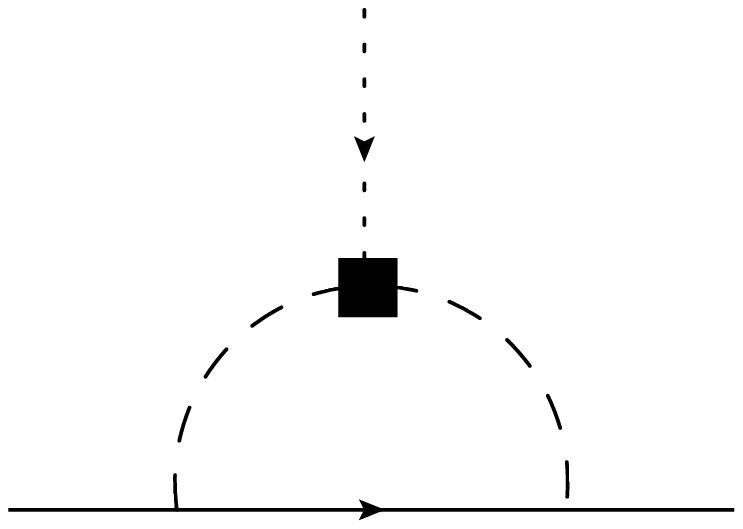}} 
\subfigure[]{
\label{fetaNloopaD}
\includegraphics[width=0.3\textwidth]{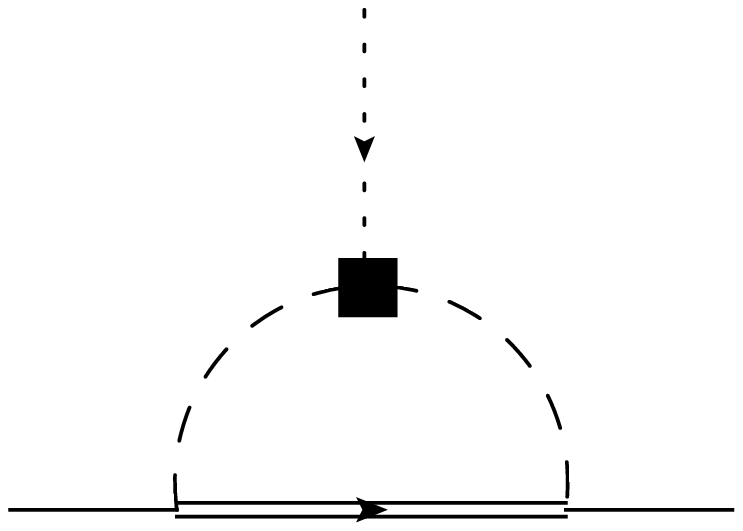}}
\end{center}
\caption{Loops that can contribute to the $CP$-violating coupling of the 
$\eta(\eta')$ 
to the nucleon. The single solid lines stand for nucleons, the double lines 
for the 
$\Delta$, the dashes for pions and the dotted lines for the $\eta(\eta')$. 
The black box at the $H\pi\pi$ vertex indicates the $CP$-violating coupling.}
\label{fetaNloop}
\end{figure}

However, the goal of this work is to set constraints on the size of the CP-violating coupling of the $\eta(\eta')$ to two pions, $f_{H\pi\pi}$. Thus we need to relate $g^{CP}_{HNN}$ to $f_{H\pi\pi}$, which is done via the ansatz shown in Fig.~\ref{fetaNloop}: there, the $\eta(\eta')$ couples to the neutron via an intermediate pion loop. These loops, that include the coupling $f_{H\pi\pi}$, when contracted to one point give precisely the coupling $g^{CP}_{HNN}$. With this approach, one can thus relate the unknown coupling $f_{H\pi\pi}$ to $g^{CP}_{H\pi\pi}$, which was extracted from the size of the neutron EDM.

The effective Lagrangian describing this 
$CP$-violating $\eta(\eta') \pi\pi$ coupling 
is given by~\cite{Gorchtein:2008pe}:
\begin{align}
\mathcal{L}^\text{CP}_{H\pi\pi}=f_{H\pi\pi} \, 
M_H \, H \, \vec{\pi}^2\,, \label{ELagCPepp}
\end{align} 
where $M_H$ is the mass of the $\eta(\eta')$ meson and 
$f_{H\pi\pi}$ is the coupling constant of $\eta(\eta')$ 
to the pions. Considering the current experimental upper limit of $2.9\cdot 10^{-26}e$~cm on 
the neutron EDM, the constraint on the CP-violating couplings $f_{H\pi\pi}$ is eight orders of magnitude lower than experimentally measured so far, as shown in Eq.~\ref{Epdgbranch}.

It is important to keep in mind that the sources considered here to originate the neutron EDM included only the CP-violating coupling of $\eta(\eta')$ mesons to the neutron. In the future, our goal is to extend this work by considering additional contributions, such as vertices where a photon couples directly to the CP-violating meson-nucleon vertex. The contributions from the additional loops that arise from this kind of vertex are expected to give even larger contributions, since they are of a lower chiral order, thus constraining the $\eta(\eta')\rightarrow\pi\pi$ branching ratio even further. These vertices can be estimated through contributions similar to those in Fig.~\ref{fetaNloop}, but with additional external photon lines, i.e.\ by pion and $\eta(\eta')$ photoproduction with one CP-violating vertex.

\section{Summary}
\label{Ssummedm} 

We related the neutron EDM to the $CP$-violating decay of the $\eta(\eta')$ mesons into two pions, with the help of fully covariant ChPT. In particular, since the experimental upper limit of $2.9\cdot 10^{-26}e$~cm for the neutron EDM is very small, it sets a very strong constraint on observables related to it. Therefore, here the goal was to give an estimate of 
the size of the CP-violating branching ratio.

In total, 
we obtained a constraint on the $CP$-violating $\eta(\eta')\rightarrow\pi\pi$ 
decay ratio roughly eight orders of magnitude smaller than measured 
in the experiment so far. This is a very instructive result, since it gives 
an estimate in a region where experimental results are not yet achievable.

It is important to keep in mind that the constraints given in this work are mere estimates, since other processes than those considered here give additional contributions to the neutron EDM. Nevertheless, due to the very large discrepancy between the experimental 
constraint on 
the CP-violating decays of the $\eta(\eta')$ and those calculated from the current upper limit for the neutron EDM, the results are still rigorous enough for the conclusions to persist, even if other processes are to be 
additionally considered.

\section*{Acknowledgements} 

This work was supported by the Spanish Ministerio de Econom\'ia y Competitividad (MINECO) and the European fund for regional development (EFRD) under contracts No.\ FIS2014-51948-C2-2-P and No.\ SEV-2014-0398. It has also been supported by Generalitat Valenciana under contract PROMETEOII/2014/0068, the Deutsche Forschungsgemeinschaft DFG, the German Bundesministerium f\"ur Bildung und Forschung (BMBF) 
under Project 05P2015 - ALICE at High Rate (BMBF-FSP 202): 
``Jet- and fragmentation processes at ALICE and the parton structure  
of nuclei and structure of heavy hadrons'', 
by CONICYT (Chile)  Basal No. FB082 and Ring No.~ACT1406, 
by FONDECYT  (Chile) Grants No.~1140471 and No.~1150792, 
by Tomsk State University Competitiveness
Improvement Program and the Russian Federation program ``Nauka''
(Contract No. 0.1764.GZB.2017), and by Tomsk Polytechnic University
Competitiveness Enhancement Program (Grant No. VIU-FTI-72/2017).


\begin{thebibliography}{99} 

\bibitem{Agashe:2014kda} 
C.~Patrignani et al. (Particle Data Group), 
\emph{Review of Particle Physics}, 
Chin.\ Phys.\ C \textbf{40}, 100001 (2016).

\bibitem{Kuckei:2005pg} 
J.~Kuckei, C.~Dib, A.~Faessler, T.~Gutsche, S.~Kovalenko, 
V.~E.~Lyubovitskij, and K.~Pumsa-ard, 
\emph{Strong CP violation and the neutron electric dipole form-factor}, 
Phys.\ Atom.\ Nucl.\ \textbf{70}, 349 (2007).

\bibitem{Pospelov:2005pr} 
M.~Pospelov and A.~Ritz, 
\emph{Electric dipole moments as probes of new physics}, 
Annals Phys.\ \textbf{318}, 119 (2005).

\bibitem{Faessler:2006at} 
A.~Faessler, T.~Gutsche, S.~Kovalenko, and V.~E.~Lyubovitskij, 
\emph{Implications of $R$-parity violating supersymmetry for atomic 
and hadronic EDMs},
Phys.\ Rev.\ D \textbf{74}, 074013 (2006).

\bibitem{Dib:2006hk} 
C.~Dib, A.~Faessler, T.~Gutsche, S.~Kovalenko, J.~Kuckei, 
V.~E.~Lyubovitskij, and K.~Pumsa-ard, 
\emph{The Neutron electric dipole form-factor in the perturbative 
chiral quark model}, 
J.\ Phys.\ G \textbf{32}, 547 (2006).

\bibitem{Faessler:2006vi} 
A.~Faessler, T.~Gutsche, S.~Kovalenko, and V.~E.~Lyubovitskij, 
\emph{Hadronic electric dipole moments in R-parity violating supersymmetry},
Phys.\ Rev.\ D \textbf{73}, 114023 (2006).

\bibitem{Pich:1991fq} 
A.~Pich and E.~de Rafael, 
\emph{Strong CP violation in an effective chiral Lagrangian approach}, 
Nucl.\ Phys.\ B \textbf{367}, 313 (1991).

\bibitem{Dar:2000tn}  
S.~Dar, \emph{The Neutron EDM in the SM: A Review}, 
hep-ph/0008248 (2000). 

\bibitem{Gorchtein:2008pe}  
M.~Gorchtein, 
\emph{Nucleon EDM and rare decays of $\eta$ and $\eta^\prime$ mesons}, 
0803.2906 [hep-ph] (2008).

\bibitem{Jarlskog:1995ww}  
C.~Jarlskog and E.~Shabalin, 
\emph{How large are the rates of the CP violating 
$\eta$, $\eta^\prime \rightarrow \pi\pi$ decays?}, 
Phys.\ Rev.\ D \textbf{52}, 248 (1995). 

\bibitem{Peccei:1977hh} 
R.~D.~Peccei and H.~R.~Quinn, 
\emph{CP Conservation in the Presence of Instantons}, 
Phys.\ Rev.\ Lett.\ \textbf{38}, 1440 (1977).

\bibitem{Peccei:1977ur} 
R.~D.~Peccei and H.~R.~Quinn, 
\emph{Constraints Imposed by CP Conservation in the Presence of Instantons}, 
Phys.\ Rev.\ D \textbf{16}, 1791 (1977).

\bibitem{Aaij:2016jaa} 
R.~Aaij et al. (LHCb Collaboration), 
\emph{Search for the $C\!P$-violating strong decays 
$\eta \to \pi^+\pi^-$ and $\eta^\prime(958) \to \pi^+\pi^-$}, 
Phys.\ Lett.\ B {\bf 764}, 233 (2017). 

\bibitem{Ghoul:2015ifw} 
H.~Al Ghoul et al. (GlueX Collaboration), 
\emph{First Results from The GlueX Experiment},
AIP Conf.\ Proc.\ \textbf{1735}, 020001 (2016).

\bibitem{Weinberg:1978kz}
 S.~Weinberg,
 \emph{Phenomenological Lagrangians},
 Physica A {\bf 96}, 327 (1979).

\bibitem{Gasser:1987rb} 
J.~Gasser, M.~E.~Sainio, and A.~Svarc, 
\emph{Nucleons with Chiral Loops}, 
Nucl.\ Phys.\ B {\bf 307}, 779 (1988).

\bibitem{Borasoy:1995ds}
 B.~Borasoy and Ulf-G.~Mei{\ss}ner,
 \emph{Chiral Lagrangians for baryons coupled to massive spin 1
                        fields},
 Int.\ J.\ Mod.\ Phys. A {\bf 11}, 5183 (1996).

\bibitem{Drechsel:1998hk}
 D.~Drechsel, O.~Hanstein, S.~S.~Kamalov, and L.~Tiator,
 \emph{A Unitary isobar model for pion photoproduction and
                        electroproduction on the proton up to 1-GeV},
 Nucl.\ Phys.\ A {\bf 645}, 145 (1999).

\bibitem{Kubis:2000aa}
 B.~Kubis and Ulf-G.~Mei{\ss}ner,
 \emph{Baryon form-factors in chiral perturbation theory},
 Eur.\ Phys.\ J.\ C {\bf 18}, 747 (2001).
 
\bibitem{Schindler:2005ke}
 M.~R.~Schindler, J.~Gegelia, and S.~Scherer,
 \emph{Electromagnetic form-factors of the nucleon in chiral
                        perturbation theory including vector mesons},
 Eur.\ Phys.\ J.\ A {\bf 26}, 1 (2005).

\bibitem{Pascalutsa:2000kd}
 V.~Pascalutsa,
 \emph{Correspondence of consistent and inconsistent spin - 3/2
                        couplings via the equivalence theorem},
 Phys.\ Lett.\ B {\bf 503}, 85 (2001).

\bibitem{Gegelia:1999gf} 
J.~Gegelia and G.~Japaridze, 
\emph{Matching heavy particle approach to relativistic theory},
Phys.\ Rev.\ D \textbf{60}, 114038 (1999).

\bibitem{Fuchs:2003qc} 
T.~Fuchs, J.~Gegelia, G.~Japaridze, and S.~Scherer, 
\emph{Renormalization of relativistic baryon chiral perturbation theory 
and power counting},
Phys.\ Rev.\ D \textbf{68}, 056005 (2003).
 
 \bibitem{Gutsche:2016jap}
  T.~Gutsche, A.~N.~Hiller Blin, S.~Kovalenko,
 S.~Kuleshov, V.~E.~Lyubovitskij, M.~J.~Vicente Vacas, and A.~Zhevlakov,
 \emph{CP-violating decays of the pseudoscalars eta and eta'
                        and their connection to the electric dipole moment of the
                        neutron},
 Phys.\ Rev.\ D {\bf 95}, 036022 (2017).

\end{thebibliography}
\end{document}